# Recent Advances in the Internet of Medical Things (IoMT) Systems Security

Ali Ghubaish, Graduate Student Member, IEEE, Tara Salman, Graduate Student Member, IEEE, Maede Zolanvari, Graduate Student Member, IEEE, Devrim Unal, Senior Member, IEEE, Abdulla Al-Ali, Member, IEEE, Raj Jain, Life Fellow, IEEE

*Abstract*— The rapid evolutions in micro-computing, mini-hardware manufacturing, and machine to machine (M2M) communications have enabled novel Internet of Things (IoT) solutions to reshape many networking applications. Healthcare systems are among these applications that have been revolutionized with IoT, introducing an IoT branch known as the Internet of Medical Things (IoMT) systems. IoMT systems allow remote monitoring of patients with chronic diseases. Thus, it can provide timely patients' diagnostic that can save their life in case of emergencies. However, security in these critical systems is a major challenge facing their wide utilization.

In this paper, we present state-of-the-art techniques to secure IoMT systems' data during collection, transmission, and storage. We comprehensively overview IoMT systems' potential attacks, including physical and network attacks. Our findings reveal that most security techniques do not consider various types of attacks. Hence, we propose a security framework that combines several security techniques. The framework covers IoMT security requirements and can mitigate most of its known attacks.

## I. Introduction

OVER recent years, Internet of Things (IoT) solutions have increasingly been adopted by many applications affecting our daily life. According to Gartner, by the end of 2020, we will have 5.8 billion IoT devices, increasing by 20% compared to last year [1]. These devices are compact, can communicate in various ways, and have low power consumption.

Internet of Medical Things (IoMT) is an IoT branch dedicated to the healthcare industry. By the end of 2020, IoMT devices are going to constitute 40% of the IoT market. This is expected to expand in the next couple of years due to the IoMT devices' potential contribution in reducing the healthcare industry expenses [2]. This industry can save up to $300 billion by relying more on IoMT devices, especially for chronic diseases and telehealth [3]. The IoMT revenue in 2017 was $28 billion and is projected to be $135 billion by 2025, making it an attractive market for investors [4].

However, the security of IoMT devices and healthcare systems as a whole (henceforth, IoMT systems) is a major challenge. The healthcare data involved in IoMT systems should be protected at various stages, including data collection, transmission, and storage. According to the 2020 CyberMDX report [5], nearly half of IoMT devices are vulnerable to exploits. IoMT systems differ from other systems since they can affect patients' lives and impose privacy concerns if patients' identities are revealed. Further, healthcare data's average cost is 50 times more than that for credit card information, making them highly valuable on the black market [6].

Hence, security is one of the main requirements for the success of IoMT systems. These systems have a set of 11 security requirements to provide data confidentiality, integrity, availability, non-repudiation, and authentication, which are referred to as CIANA [7]. These requirements can be satisfied by traditional security solutions also. However, due to their power consumption and other system requirements, traditional solutions may fail to give proper security guarantees. Researchers have instead proposed several techniques that are designed explicitly for IoMT and IoT systems. These techniques can be divided into symmetric cryptography, asymmetric cryptography, and keyless non-cryptographic techniques based on cryptography.

Most of the review literature on IoMT systems discuss their limitations, security issues, and solutions. For example, Yaacoub *et al.* classify the security techniques in IoMT systems and wireless body area networks (WBANs) in general into cryptographic and non-cryptographic [8]. They categorize the countermeasures into authorization, availability, intrusion detection systems (IDSs), and awareness. Vyas and Pal discuss the open issues in these networks, such as flexibility, single point of failure, and handling emergencies [9]. The design and security challenges, such as securing patient data in the cloud for wearable devices in IoMT systems, have been discussed by Bhushan and Agrawal [10].

Machine learning (ML), artificial intelligence (AI), and the blockchain technology are other techniques considered to secure IoMT systems [11, 12]. These techniques can enhance

This paper has been submitted on Aug. 30th, 2020 to IEEE Internet of Things Journal Special Issue on Internet of Things for Smart Health and Emotion Care. This work was supported in part by the National Priorities Research Program (NPRP) from the Qatar National Research Fund (QNRF) under Award NPRP-10-0125-170250 (a member of the Qatar Foundation), in part by the NSF under Grant CNS-1718929, and in part by Prince Sattam Bin Abdulaziz University, Al-Kharj, Saudi Arabia.

Ali Ghubaish, Tara Salman, Maede Zolanvari, and Raj Jain are with the Department of Computer Science and Engineering, Washington University in Saint Louis, St. Louis, MO 63130 USA (e-mail: aghubaish@wustl.edu; tara.salman@wustl.edu; maede.zolanvari@wustl.edu; jain@wustl.edu).

Devrim Unal is with the KINDI Center for Computing Research, College of Engineering, Qatar University, Doha, Qatar (email: dunal@qu.edu.qa).

Abdulla Al-Ali is with the Computer Science and Engineering Department, College of Engineering, Qatar University, Doha, Qatar (email: abdulla.alali@qu.edu.qa).

the systems' performance and provide tolerance to some attacks and issues like denial-of-service (DoS) attacks and a single point of failure. Also, ML techniques can reduce the physical layer authentication error by 64% compared to traditional authentication methods [13, 14].

With the advancements in both security protection techniques and new types of attacks targeting IoMT systems, a full review of current IoMT systems' security techniques and attacks is required. Therefore, this paper reviews state-of-the-art security and attack techniques for IoMT systems. Our main contributions can be summarized as follows:
1) We review the security requirements that are necessary for IoMT systems as well as different types of techniques to provide secure data collection, transmission, and storage.
2) We discuss the available security techniques and their resilience against different types of attacks. We argue that any single technique cannot provide comprehensive security against most known attacks targeting these systems.
3) We explore the IoMT attack surface and show the resilience of these security techniques against these attacks. This includes new attacks that have recently targeted IoMT systems.
4) We propose a security framework that uses some of the features of these techniques for IoMT systems. This framework covers IoMT systems' security during data collection, transmission, and storage. We have considered the constraints for these devices in our framework.

The rest of the paper is organized as follows: A brief background of the IoMT systems types and architecture is provided in Section II. In Section III, we present IoMT threats at different stages, along with security requirements and different types of security techniques. State-of-the-art security techniques, including symmetric, asymmetric, and keyless, are discussed in Section IV, V, and VI, respectively. The IoMT attack surface is described in Section VII, while our proposed security framework is presented in Section VIII. Finally, we conclude the paper in Section IX.

## II. Background

This section provides a background on IoMT systems as well as their architecture. This helps to understand the later sections, where we present IoMT systems' security requirements, attacks, and countermeasures.

### A. IoMT Types

IoMT systems provide the necessary or improved assistance for many medical conditions. The necessary devices are implantable devices for particular medical conditions, e.g., pacemakers for heart conditions. On the other hand, the assisting devices are mostly wearables for improved healthcare experience, e.g., smartwatches. These differences put the IoMT systems into two categories, implantable medical devices (IMDs) and Internet of wearable devices (IoWDs).
1) *Implantable Medical Devices (IMDs)*

Any device that is implanted to replace, support, or enhance a biological structure is an IMD. For example, a pacemaker is an IMD that helps control abnormal heart rhythms, i.e., by promoting the heart to beat at a normal rate if it is beating too fast or too slow [15]. Fig. 1 shows several popular IMDs and their placement locations in the human body. Recently, wireless IMDs have been proposed to solve problems associated with wired IMDs, e.g., infection and cable breakage [16]. IMDs are mostly very small and have very long battery life. Hence, low power consumption, small storage space, and small batteries that last long are essential requirements for these devices to stay inside a human body for a long time. For instance, pacemaker implants tend to last 5 to 15 years [17].

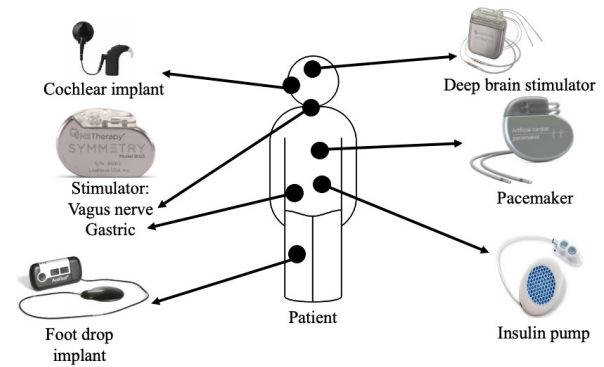

Fig. 1. Examples of IMDs and their locations in the human body.

2) *Internet of Wearable Devices (IoWDs)*

These are devices worn by individuals to monitor their biometrics, e.g., heart rate, and may help improve individuals' overall health. Examples include smartwatches, fall detection band, electrocardiogram (ECG) monitors, and blood pressure monitors [18]. Smartwatches are currently one of the most known forms of IoWDs to monitor biometrics such as heart rate and movement. The monitoring can be used to detect slow and fast heartbeats when the individual is not active. The new watches also support fall detection and ECG readings to detect atrial fibrillation (irregular heartbeat) medical conditions. They are currently widely used for non-critical patient monitoring [19]. However, these devices have sensor accuracy and battery life limitations; thus, not likely to replace IMDs in critical conditions [20].

### B. IoMT Systems Architecture

Most of the current IoMT systems are typically divided into four layers, as shown in Fig. 2 [21]. These layers include all data stages starting from the individual's biometric collection stage and ending in data storage and subsequent visualization by a physician for analysis. Moreover, the patient can also visualize their overall health status from the cloud. With the current advances in IMDs, IoWDs, and IMDs mostly share the same architecture given that IMDs can communicate with the gateways, as exemplified by Medtronic peacemaker [22].
1) *Sensor Layer*

This layer consists of a set of small implanted or worn sensors that collect the patient's biometrics. The data are transmitted to the second layer over wireless protocols such as Wi-Fi, Bluetooth, or over MedRadio frequency spectrum reserved for IMDs [23].

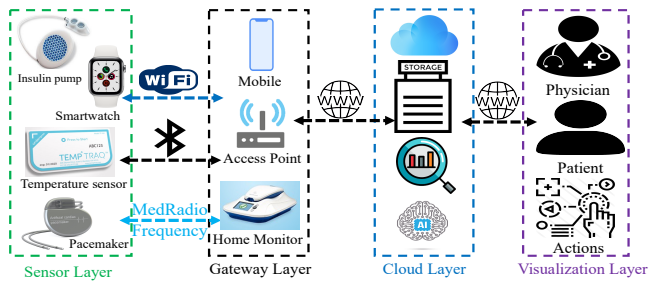

Fig. 2. IoMT system architecture.

2) *Gateway Layer*

Due to the processing and storage limitations of IoMT sensors, the data are transferred without processing to the second layer, i.e., the gateway layer. The devices in this layer can be the patient's smartphone or a dedicated access point (AP), which are generally more powerful than sensors. They can perform some preprocessing operations, such as validation, short term data storage, and simple AI-based analysis. In addition, they send the sensor data to the cloud over the Internet.

3) *Cloud Layer*

The cloud layer is responsible for getting the data from the gateway for storage, analysis, and secure access. The analysis may include data processing to find any changes in the patient's health and presenting them to the physicians or patients for further actions. The key generation server (KGS) is responsible for generating IDs and keys for various system nodes. The access to the sensors can be remotely managed and controlled from this layer.

4) *Visualization/Action Layer*

In this layer, the data are presented to the physicians and the patients to track their health. This layer also includes the actions recommended by the physician based on the patient's health conditions. Examples of actions include prescribing or adjusting the dosage for various medicines.

## III. IoMT Security Model

In this section, we discuss the threats to the IoMT systems' data at three different stages. Also, we present the IoMT systems' security requirements and generally categorize countermeasure techniques. In subsequent sections, each countermeasure category will be further detailed with its associated techniques and use in IoMT systems.

### A. IoMT Threats at Different Stages

IoMT systems must protect the patients' data at all stages, including collection, transmission, and storage. As shown in Fig. 2, these stages consist of combinations of the four architecture layers.

1) *Data Collection*

The collection of the patient's data in the sensor layer is the first stage of an IoMT system. Attacks at this stage can be software (i.e., data tampering) or hardware (i.e., sensor hardware manipulation) attacks. These attacks can threaten patients' life if the sensor hardware or software is affected. Thus, protecting the data against these attacks is vital to keep the system running.

2) *Data in Transit*

This stage includes communications between the devices in all four layers, e.g., the communications between the IoMT sensors in the sensor layer and the AP in the gateway layers. Attacks here can manipulate or block the sensor data being transmitted. Thus, securing against these attacks would prevent the data from being affected while being transferred among the four layers.

3) *Data in Storage*

After the patient's data are collected and transmitted from the sensor and gateway layers, they are stored in the cloud layer. Attacks in this layer vary from stealing account credentials to DoS or distributed DoS (DDoS) attacks. Protecting the data in this layer and the visualization layer from any unauthorized access is essential. This is critical since, in this layer, most of the data are resting most of the time; hence, they are at more risk than any other stage.

### B. IoMT Security Requirements

Due to the patient data's sensitivity and safety, a set of requirements that can ensure IoMT systems' security at all layers is needed. The set has been derived from CIANA considerations and consists of the following 11 security requirements [24, 25]:

1) *Confidentiality/Privacy*

The ability to keep the data private while being gathered, transmitted, or stored. In addition, they must only be accessible to authorized users. The most common techniques to fulfill this requirement are data encryption and access control lists, which will be discussed further in the next section.

2) *Integrity*

This is related to the capability of protecting the data from any unauthorized tampering during the collection, transmission, and storage stages.

3) *Availability*

The ability to correctly keep the IoMT systems continuously running. This can be done by keeping the system up to date, monitoring any changes in their performance, providing redundant data storage or transmission routes in case of DoS attacks, and fixing any problem as soon as possible.

4) *Non-Repudiation*

The ability to make each authorized user responsible for his/her actions. In other words, this requirement guarantees that any interaction in the system cannot be denied. This can be achieved using digital signature techniques, as will be discussed later in the paper.

5) *Authentication*

The capability to validate the identity of a user accessing the system. Mutual authentication is the most secure form where both the server and the client authenticate each other before any secure data/key exchange.

6) *Authorization*

The ability to allow authenticated users to only execute commands to which they are authorized. Similar to confidentiality, authorization can be achieved using proper data encryption and access control techniques.

7) *Anonymity*

The capability to keep the patients'/physicians' identities hidden from unauthorized users when they interact with the system. Using smart cards can fulfill the anonymity requirement.

8) *Forward/Backward Secrecy*

Forward secrecy provides the ability to keep future transmitted data/keys safe even if old data/keys are compromised. Backward secrecy ensures the opposite, where old data/keys are safe even if an attack has successfully affected current data/keys. Forward/Backward secrecy can be achieved by time-based authentication parameters, e.g., time-based keys that can be generated and used only when clock time at both nodes match.

9) *Secure Key Exchange*

The ability to securely share the keys between the nodes in the system. Diffie-Hellman key exchange is an example of a secure key exchange.

10) *Key-Escrow Resilience*

The system administrator cannot impersonate any authorized user in the system. This protects against internal threats. Using asymmetric keys with a cryptographic hash function (CHF) can fulfill this requirement.

11) *Session Key Agreement*

The nodes in the system must use session keys after the authentication process. Similar to key-escrow resilience, using symmetric/asymmetric keys with CHF can fulfill this requirement.

C. *IoMT Systems Security Techniques*

There are several different techniques to secure IoMT systems. These techniques can be divided into three main categories: symmetric, asymmetric, and keyless, as shown in Fig. 3. Symmetric and asymmetric techniques rely on cryptographic algorithms, while keyless techniques are non-cryptographic. The cryptographic techniques include one-factor and two-factor authentication methods, which are explained in the next three sections. One-factor authentication uses only one authentication technique to protect the system. In contrast, two-factor authentication adds a second authentication technique (factor), such as biometrics, to protect the system if one of the two factors is compromised.

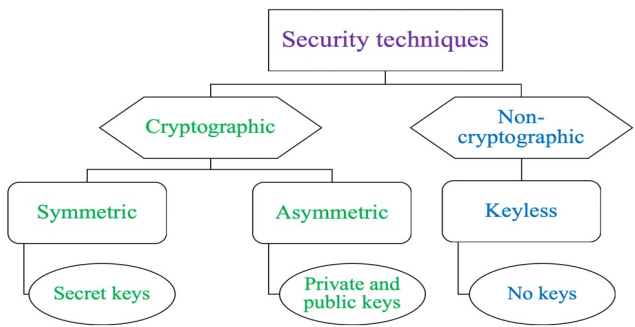

Fig. 3. Security techniques.

IV. SYMMETRIC-KEY ALGORITHMS

As shown in Fig. 4, symmetric cryptography includes any cryptographic algorithm based on a secret/shared key between two or more nodes wanting to communicate. The key is to be generated and distributed prior to using asymmetric cryptography or a prior communication stage.

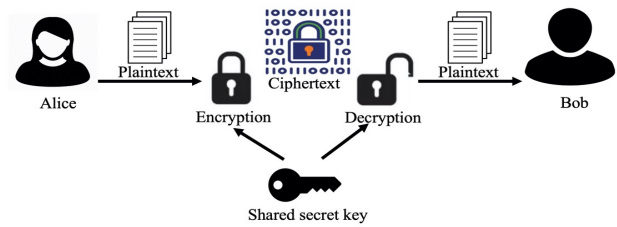

Fig. 4. Symmetric cryptography.

In this section, we discuss the integration of symmetric cryptographic algorithms in IoMT systems. These algorithms can be used for IoMT systems to allow hierarchical access to the patient's data and initiate secure connections without prior setup. Further, they can also be used in two-factor authentication where they act as a first factor while other techniques, such as facial recognition, and pattern-based act as a second factor.

A. *Hierarchical Access*

This technique allows hierarchical access control to patients' data stored in the cloud layer. One approach utilizes a hierarchical role-based model and provides authorization based on the user's role [25]. For example, all authenticated nurses can administer medicines, but prescribing a new medication requires a person authenticated as a doctor. The model supports a relatively low complex hierarchical security scheme that encrypts the patients' data and only decrypts that part of the data to which the user is authorized. Belkhouja et al. [25], have used the Chinese remainder theorem (CRT) to support this hierarchal access where the user with a higher privilege can access any patient's data. In contrast, the user with a lower privilege can access part of the data related to their roles [26].

B. *Wireless Signal Characteristics*

This technique utilizes wireless signal characteristics to secure IoMT systems by generating keys without prior connections. The radio signal strength (RSS) is one of these characteristics, and it measures the received signal power, which varies based on the medium it passes through [27]. IMDs can be excellent candidates for this technique since the RSS value variation inside the human body is different from outside the body [28]. The proposed technique uses the randomness in RSS values to generate a shared key. This key can be used to secure the communication between a headless cardiac pacemaker and a subcutaneous (under-the-skin) implant without prior knowledge of the keys. In this technique, two bits can be extracted from a single cardiac cycle (a beat) with a 128-bit key in 60 seconds if we consider the average human heart rate of 64 beats per minute (bpm).

C. *CHF with XOR*

CHF is a one-way mathematical function that converts an arbitrary size data to a fixed size [29]. Exclusive-OR (XOR) can be used to check if one of its operands is different. In a medical setting, initial parameters (i.e., a sensor ID and a shared key) can be XORed together and then hashed. Then, these hashed parameters are shared from the key generation server to the sensor and gateway nodes. These nodes can generate their keys with the help of these parameters [30]. Combining the CHF, a symmetric key, and the XOR operator can secure the IoMT systems' communications using new authenticated key agreement protocols, as illustrated by Alzahrani et al. [31] and Xu et al. [31, 32]. This technique also supports unique identification parameters for the system's nodes using the hash function. However, the system administrator must manually add initial parameters to all the nodes in the system's initialization step.

### D. Gait-Based Technique

This technique uses the human walking pattern to generate unique symmetric keys. A system proposed by Sun and Lo can generate a symmetric key using a set of IoMT sensors attached to the individual's body in just a matter of 10 gait-cycles. They claim that their system can generate three times the number of bits per gait cycle than those generated by similar state-of-the-art techniques [33]. The gait cycle is defined as one cycle of movement between two repetitive events while walking. This system employs an artificial neural network (ANN) model to generate 13 b/gait-cycle, which will generate a 128-bit key in just a matter of 10 gait-cycles. This key can be used later to secure the communications between the IoMT sensors and the AP or mobile in the gateway layer. It outperforms finger-based systems by generating binary keys at different times, which provides randomness to the keys without direct user interaction with the system.

### E. Facial Recognition

This technology is a one-way that IoMT systems can rely on authenticating users by scanning their faces. By using shared keys as a first factor, facial recognition can be used as a second factor in continuous role-based authentication [34]. This helps keep the connection between the sensor and the medical controller in the gateway layer secure and based on each authorized user's privilege. Since this technique is continuously scanning the user's face while using the system, it can secure the system in a medical setting. For example, this technique can prevent the medical staff with lower privileges from accessing the patient's data in the absence of a higher privileged medical staff that has authenticated himself/herself but has not logged out from the system.

### F. Pattern-Based Technique

This technique is similar to the facial recognition system, but it uses a pattern-based technique as a second factor [35]. This technique uses a tab pattern generated to be performed by the patient to control the sensor. After successfully passing the first factor with the medical controller in the gateway layer, the controller sends a random tab pattern as a second factor to the user before executing a sensitive command. The technique can also be used to keep the sensor communication turned off until a specific pattern is performed, preserving the sensor's battery power in case of IMDs.

## V. Asymmetric-key Algorithms

Asymmetric cryptography includes cryptographic algorithms that use two keys, a public and a private, with one of them for encryption/validation, and the other is used for decryption/signature. Asymmetric cryptography is also known as public-key cryptography. The public key is known to everyone, while the private key is only known to its owner. An example of how encryption and decryption can be used is shown in Fig. 5. Some of the known algorithms in this category include Rivest–Shamir–Adleman (RSA) and Elliptic-Curve Cryptography (ECC) [36, 37]. ECC is the most common encryption technique used for securing IoMT systems due to its lightweight characteristics. An ECC key with a size of 160 bits is as good as the 1024-bit RSA key and is 15 times faster [38].

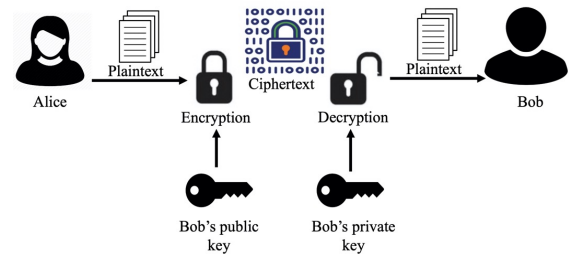

Fig. 5. Asymmetric cryptography.

This section discusses the integration of asymmetric cryptographic algorithms in IoMT systems. This includes asymmetric keys with CHF, homographic encryption (HE), or digital signatures. Also, similar to symmetric keys, asymmetric keys can be used for two-factor authentication. They act as a first factor for authentication with other techniques, such as smart cards as a second factor. Smart cards are extensively used in hospitals nowadays.

### A. CHF with ECC

CHF function, along with ECC keys, can be used as a secure certificateless channel between the patients and their medical doctors [24]. The idea of combining the ECC and the CHF is to allow a secure way for sharing keys between the key generation server in the cloud layer and the nodes in the IoMT sensor and gateway layers, respectively. The ECC public key of the KGS and initial parameters, such as a node ID, are hashed together using CHF; then, they are sent to the nodes in the IoMT sensor and gateway layers. The nodes can generate their asymmetric keys with the help of the received hashed values. As a result, this system solves the problem of sharing the secret keys as in the symmetric cryptographic techniques.

It can also overcome the overhead in certificate management for data storage and sharing in the cloud [39]. The IoMT data sizes are substantial, and they are increasing. We are in the zettabyte era. The zettabyte is one billion terabytes, where a terabyte is a typical hard disk size nowadays [40]. By dividing the patient's data into subsets and converting them using ECC keys and CHF, they can be securely shared among the system's entities. The average energy consumption in this technique is around 30% less than similar techniques.

### B. Homomorphic Encryption (HE)

HE is an encryption technique that preserves data confidentiality and allows limited mathematical operations to be done on encrypted data [41]. This technique protects the patient's data privacy and stores them as ciphertext in the cloud layer to do mathematical operations, such as data integrity.

However, this technique differs from other techniques since it allows only the patient to see their data and not the medical staff except during emergencies. In other words, this is useful for some IoMT sensors, such as a smartwatch, which allows the data to be encrypted at all times and only seen by the patient except in emergencies where the patient's data can be sent to the medical staff to make correct diagnostics.

There are three different schemes for HE: partial HE (PHE), somewhat HE (SHE), and fully HE (FHE). PHE supports one mathematical operation for an unlimited number of times, while SHE supports only a limited number of operations. FHE supports an unlimited number of operations, and therefore, it can be suitable for fast aggregation of data without compromising data confidentiality [38]. Hence, it is ideal for healthcare monitoring systems in hospitals. Jariwala and Jinwala claim that their AdaptableSDA HE framework consumes only 10% more power with the privacy requirement than without it [38].

Optimal HE (OHE) is a modification of FHE. It differs from FHE in that it is based on the Step-size Firefly Optimization (SFFO) algorithm in which the key with the maximum breaking time is selected [42]. This technique reduces the computation time and increases the breaking time by 2% to 8% compared to other HE and non-HE techniques individually.

### C. Digital Signatures

Digital signature techniques can be used even in a small IoMT system. In general, they can be used to verify the data/command authenticity using the sender's (Alice) private and public keys for signature and verification, respectively [43]. In IoMT systems, digital signatures can be integrated into the sensor's firmware with an add-on software shim, intercepting and validating the sensor's wireless communications [44]. These techniques require storing a list of authorized users' (i.e., medical staff's) public keys in the sensor's firmware to validate them.

### D. Smart Cards

This technique is different from the first three techniques since it relies on physical keys [45]. These keys are utilized as a second factor, with the ECC keys as the first factor for authentication. In IoMT settings, the medical staff must first enter a key and then use their smart cards to access the system. This technique helps the system be resistant to cyber-breaks if one of the factors is stolen or lost. This has made them quite common nowadays.

## VI. KEYLESS ALGORITHMS

In this section, we discuss keyless techniques that provide security without pre-shared keys. The techniques in this category can be based on biometrics, token-based security, or proxy-based techniques. Cutting edge technologies such as the blockchain technology and AI also fall in this category since they can be used for security without pre-shared keys.

### A. Biometrics

The biometric sensors used to identify users' physical characteristics are the most common technique employed to provide security for IoMT systems since they are easy to use. In a medical setting, either the medical staff or the patient can access the medical records by only using their biometrics. Biometric factors include fingerprint and ECG-based sensors that are handy in case of emergencies. The fingerprint sensors are based on reading the fingerprint image, while the ECG-based sensors record the heartbeat activities to encrypt the data. Fingerprint sensors reduce the messages' size during transmission and the computational overhead compared to the ECG-based techniques [46].

The performance of the fingerprint sensors is based on the extraction algorithm that is used. Popular algorithms used in these sensors are Delaunay Triangulation-based feature representation, Pair-polar coordinate-based feature representation, and Minutia Cylinder-Code-based feature representation [47]. According to Zheng et al., Delaunay shows better performance and is less complicated than the other techniques. The advantages of using fingerprint biometrics include their long history and credibility than face recognition-based systems.

### B. Token-Based Security

User authentication can be done using software or hardware tokens. For instance, the x-auth-token field in the hypertext transfer protocol (HTTP) header can be used as a software token embedded in the user web browsers [48]. Cloud data analytics companies use these tokens, e.g., IoT Ubidots [49], to secure the connection between the cloud layer and the nodes in the IoMT sensor and gateway layers. Likewise, RFID can be used as a hardware token for secure logistic management of sensors in a hospital information system (HIS) [50].

### C. Proxy-Based and Light-Based Systems

Proxy-based systems are basically made of a middleware device that controls the communication between the sensors and any device communicating with them, such as medical controllers. Besides, they can provide full-duplex secure communications between these devices, where they can simultaneously communicate. These middleware devices can be a set of microprocessors inside a jacket or a belt to be worn by the patient [51, 52].

Light-based communication technologies, such as Light-Fidelity (Li-Fi), can be used to secure the monitoring capabilities for HIS, as presented by Mosaif and Rakrak [53]. Since Li-Fi does not use wireless communications, it has no interference with the hospital network, substantial free operation frequency, and short coverage range for enhanced security.

### D. Blockchain Technology and AI

These are new techniques for use in IoMT systems due to their success in providing security in other fields, such as finance [54-56]. The blockchain technology is typically used in IoMT systems as a security management sharing technique for the data between the patient and other parties such as doctors and insurance companies. On the other hand, AI systems can detect anomaly behaviors (leading to attacks) in network flows and patients' data. However, there are some challenges for these techniques to be adopted by IoMT systems. For example, the blockchain technology may suffer from latency, storage issues, and communications overhead given the data sizes and communication requirements in IoMT systems. High latency is typical for public blockchain technology due to its distributed nature and decentralization. Therefore, private blockchains may be considered for real-time systems. AI systems require a large amount of data; hence, they may not be ideal for detecting rare attacks. Nonetheless, the blockchain technology and AI are being adopted in IoMT systems, mainly in the cloud layer [57, 58].

## VII. IoMT Systems Risks and List of Attacks

In this section, we explore the attack surface of IoMT systems. We discuss possible attacks that can target such systems, including physical and network attacks. In Table I, we summarize the security requirements for IoMT systems, possible attacks, and countermeasures [12]. As shown in the table, the countermeasures for 11 out of the 14 attacks are based on keyless methods, and more than half of all countermeasures are based on two-factor authentication methods. The popularity of these methods is due to their simplicity during system implementation and management.

### A. Physical Attacks

These attacks target the physical components (e.g., sensors, physical keys) of the IoMT systems to extract patient data or security keys. They require some component of the IoMT systems to be physically accessible to the attacker. These attacks can be summarized as follows:

1) *Physical Security Token Loss*

This includes any attack where the attacker steals an authorized user's physical security token, such as a smart card to access the system. The violated security requirements here are authentication, authorization, anonymity, and forward secrecy. Kumari *et al.* showed that integrating asymmetric keys, such as ECC with smart cards, can mitigate such attacks since stealing the smart card is insufficient to hijack the system [45].

2) *Impersonation/Presentation*

In this attack, the attacker impersonates an authorized user's identity, e.g., by replicating the fingerprint or face print. This can target any node in the IoMT system. The attack has security violations to authentication, authorization, anonymity, and forward secrecy security requirements. It can be avoided using symmetric/asymmetric techniques, such as CHF, or keyless techniques, such as biometrics [24, 31, 46, 47].

3) *Tampering*

Any type of modification to the IoMT systems' data at the collection, transit, or storage stage is considered a tampering attack. This may include attaching external devices to alter the data and attack sensors during emergencies. It violates data confidentiality and integrity and can be mitigated by combining symmetric keys with facial recognition or using keyless methods [31, 34, 39, 46].

4) *Side Channel*

These attacks occur during the communications among devices in the IoMT system. They are based on leaked information about the cryptographic operation in the communications. Data confidentiality and privacy requirements are violated by these attacks and can be alleviated using keyless cryptography. Maji *et al.* suggest using the datagram transport layer security (DTLS) protocol to avoid them. Blockchain technology and AI can act as other detection and mitigation strategies, as shown by Saif *et al.* [35, 54].

5) *Radio Frequency (RF) Jamming/Desynchronization*

RF Jamming attacks target the system's availability, which is dangerous for critical systems such as IoMT systems. Also, they can cause battery depletion, knowing that IoMT sensors are battery-power constrained. The blockchain technology and AI can reduce the effects of such intrusions by finding alternative routes or terminating the channel connection with the attacker [55].

TABLE I
LIST OF ATTACKS AND COUNTERMEASURES

| # | Attack | Effects | Countermeasure | Ref. |
|---|---|---|---|---|
| 1 | Physical security token loss | • Authentication<br>• Authorization<br>• Anonymity<br>• Forward secrecy | • Asymmetric (two-factor) | [45] |
| 2 | Impersonation | | • Asymmetric<br>• Keyless | [24, 31, 46, 47] |
| 3 | Tampering | • Data confidentiality<br>• Data Integrity | • Symmetric (two-factor)<br>• Keyless | [31, 34, 39, 46] |
| 4 | Side channel | | • Keyless | [35, 54] |
| 5 | RF jamming | • Availability | | [55] |
| 6 | DoS/DDoS | | | |
| 7 | Sniffing | • Data confidentiality | • Symmetric /asymmetric (two-factor)<br>• Keyless | [54, 59] |
| 8 | MITM | • Data confidentiality<br>• Authorization | | [24, 25] |
| 9 | Relay | • Authorization | | [25, 55] |
| 10 | Replay | | | [32, 34, 45, 54] |
| 11 | Clock synchronization | • Secure key exchange | • Asymmetric (two-factor) | [45] |
| 12 | Parallel session | • Authentication<br>• Authorization | | |
| 13 | Brute force | | | [33] |
| 14 | Stepping stone | | • Keyless | - |

### B. Network Attacks

Other attacks may target the communication between different layers of the IoMT system, such as Bluetooth or Internet links presented in Fig. 2. These attacks usually aim to steal or fabricate patients' data or block the connections between the IoMT systems' layers.

6) *DoS/ DDoS*

These attacks load the system's communication links with many undesirable connections making it unavailable for regular connections. They may also cause network fragmentation. Thus, a fragmentation attack is a special type of DDoS [60]. These attacks usually target the cloud layer in the IoMT systems to prevent the system from being available to the users (i.e., patients and medical staff); hence, it violates the availability requirement. The blockchain technology and AI can reduce the effects of such intrusions by finding alternative routes, or terminating the channel connection with the attacker [55], which are similar to those mentioned in the RF jamming attacks.

7) *Sniffing*

A sniffing attack passively intercepts the data transmitted between two nodes, resulting in patient data confidentiality violation. In a medical setting, an attacker can see the data transmitted between the layers in the IoMT system architecture, which violates the data confidentiality security requirement. Any encryption algorithm, i.e., symmetric, asymmetric, or keyless, can be used to mitigate these attacks [54, 59].

8) *Man-In-The-Middle*

MITM attack is a type of eavesdropping attack. After a successful sniffing attack, the attacker can alter the intercepted data before sending them to the original destination. For example, the attacker can change the patient's biometric data transmitted from any two layers in the IoMT system (i.e., from the sensor layer to the gateway layer). This can be done using unmanned aerial vehicles (UAV), resulting in a drone-in-the-middle (DitM) attack, as discussed by Sethuraman *et al.* [59]. To make this attack more powerful, the UAV can be connected to a cloud to perform more intensive computation in a short period of time. This attack violates authorization in addition to data confidentiality requirements and can be mitigated using encryption or two-factor authentication techniques [24, 25].

9) *Relay*

After a successful sniffing attack, the attacker can relay the intercepted data to a third node without altering them. For instance, sending the patient's data after intercepting them (i.e., from the sensor layer) to the attacker's computer before sending them to the intended layer (i.e., gateway layer). This attack breaches the authorization requirement and can be mitigated using asymmetric keys, such as hierarchal access, supporting secure session keys [25, 55].

10) *Replay*

After a successful sniffing attack, the attacker can resend the intercepted data later to the original destination without altering them. By repeating this process, this attack may also result in a DoS/DDoS attack. This attack violates the authorization requirement, which is similar to the replay attack. It can be mitigated using a timestamp, which is part of some symmetric, asymmetric, and keyless techniques [32, 34, 45, 54].

11) *Clock Synchronization*

This type of attack targets the clock synchronization protocol, which is necessary for real-time systems, such as IoMT systems. The attack violates the secure key exchange requirements. The attacker successfully initiating this attack can make relay, replay, and MITM attacks not easily detectable. This attack can be mitigated by using two-factor techniques, such as ECC with smart cards [45].

12) *Parallel Session*

These attacks break one-way authentication protocols that use asymmetric keys. The effects of such attacks are authentication and authorization violations, which can be avoided by using two-factor techniques, such as ECC with smart cards [45].

13) *Brute Force*

The attacker in this type of attack tries many credentials until successful. One way is the so-called dictionary attack, which relies on known passwords or words in dictionaries. These attacks can also be performed in the off-line phase after capturing the encrypted data decrypted with powerful machines. A dictionary attack is one of the significant problems for IoT devices since their short, simple, or factory-set default passwords can be guessed using a simple python script, making them easier to find online [61]; therefore, IoMT systems can be affected. These attacks have violated authentication and authorization security requirements as the parallel session attacks but can be alleviated using keyless methods, such as biometrics [33].

14) *Stepping Stone*

Instead of relying on one computer/host to attack the IoMT system, a chain of hosts can be used to attack the system. Sethuraman *et al.* perform this attack using a series of UAVs to extend the communication link between the UAVs and the attacker computer. Hence, The attacker can launch an attack in restricted areas (i.e., in a hospital) that are not directly accessible by the attacker [59]. This attack violates the authentication and authorization security requirements, but it can be avoided using keyless methods, such as AI.

VIII. PROPOSED SECURITY FRAMEWORK FOR IOMT

As seen in the previous section, no single technique can provide a secure environment for IoMT systems. Hence, we propose a framework capable of protecting IoMT systems from the 14 attacks mentioned in the previous section. The framework also fulfills all the security requirements required by IoMT systems. There are three parts of the framework based on the IoMT security model stages mentioned in Section III.A, as shown in Fig. 6.

*A. Securing Data Collection*

The first step in securing IoMT systems is to secure how other systems interact with them, which protects the patient's data collection stage. Two-way factor authentication techniques are good options to provide such security and resistance to some of the attacks mentioned. If one of the two factors is compromised, the other can still provide essential overall security. ECC keys are commonly used techniques as the first-factor authentication due to their lightweight keys and reliable protection [24, 39, 45].

Adopting the hierarchical access technique with ECC is a perfect way to secure data sharing to other medical staff based on their role, which has been used for other fields like smart homes [62]. This technique requires KGS, located in the cloud layer, as shown in Fig. 7. Biometric sensors are considered the most common way nowadays as a second factor due to their convenience for everyday use and emergencies [46]. These sensors are used to authenticate the patient to access the sensor layer nodes, as shown in Fig. 7. As explained in Section VI.C, proxy-based techniques can be used to provide security to existing unsecured sensors at the sensor layer [51, 52].

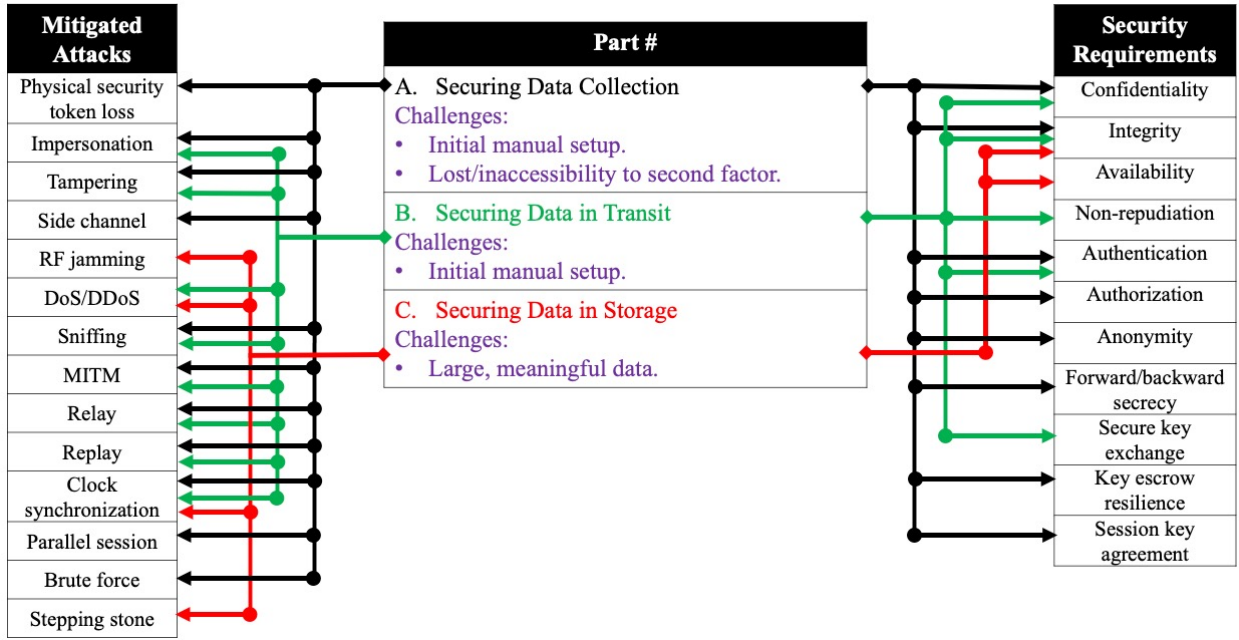

Fig. 6. Proposed framework security features.

ECC and Biometrics can protect the system in case of a software attack during the data collection stage. However, in the case of a hardware attack, the system needs another technique to alert the patient and the medical staff to reduce or eliminate the effects of such an attack. The AP or a similar device in the gateway layer should alert the user and the physician in case of not being able to connect to the IoMT sensor for a specific period of time (i.e., one hour).

Edge Computing (EC) has recently gained attention in the IoMT systems since it reduces the latency and provides powerful resources for these systems' sensors [63, 64]. EC, which is usually located in the gateway layer, as shown in Fig. 7, can act as the gateway to the IoMT sensors or as a main gateway for a set of secondary gateways. Also, it can be used here to utilize an AI model, which will be detailed in Subsection C. This model can be used to track the changes in the sensors' readings as initial analysis to fulfill the patient's data confidentiality and integrity security requirements. In case either requirement is violated, the EC can send an early warning to the patient about this violation. In case the patient did not respond, the system can immediately send an alert to his/her physician.

These techniques can provide the system with confidentiality, integrity, authentication, authorization, anonymity, forward/backward secrecy, key-escrow resilience, and session-key agreement. The system can be resilient to attacks by guaranteeing these requirements, including physical-security token, impersonation, tampering, side channel, sniffing, MITM, relay, replay, clock synchronization, parallel session, and brute force.

However, the techniques in this subsection assume pre-shared keys or initial parameters, which may lead to the following challenges:
- It requires an initial manual setup to prepare the KGS for the hierarchical access technique.
- Unusable if the second factor is lost or not accessible, especially during emergencies.

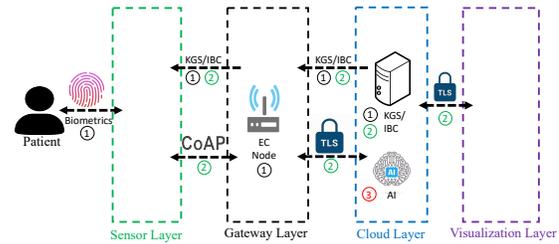

Fig. 7. Proposed IoMT secure system architecture.

B. Securing Data in Transit

To enhance the IoMT systems' security when connected to other devices over the network, we advise utilizing some security protocols, such as Constrained Application Protocol (CoAP) [65]. CoAP is an application protocol designed explicitly for resource-constrained IoT applications, such as IoMT systems, for communications between the sensor and gateway layers, as shown in Fig. 7. The rest of the layers can be linked using the Secure HTTP (HTTPs) or Transport Layer Security (TLS) version 1.3 [66]. Thus, it is convenient for use in IoMT systems. To reduce the certificate management overhead in the cloud layer, certificateless cryptography, a branch of the ID-Based Cryptography (IBC), can be used, as shown in Fig. 7 [67, 68]. The key generation process in certificateless cryptography is done using the KGS public key with some initial parameters to help the IoMT systems' nodes generate their keys. Then, Certificateless Authenticated Encryption (CLAE), which does not require central key management, is used for authentication [69]. CoAP protocol and IBC help protect the system against impersonation, tampering, sniffing, MITM, relay, replay, and clock-synchronization attacks. The protection from these attacks provides the system with confidentiality, integrity, non-repudiation, authentication, and secure key exchange. However, the system requires an initial manual setup similar to that described in Subsection VIII.A.

*C. Securing Data in Storage*

Some of the attacks in IoMT systems target the availability and integrity of the system, such as DoS/DDoS, RF jamming, and stepping-stone attacks. These attacks can be detected using AI techniques. AI techniques can be used to build detection models with mitigation techniques imposed on top of these models. For example, Deep Neural Networks (DNN) can be used to build intrusion detection models. Once this model detects suspicious activity, termination of the compromised connection is imposed to mitigate the attack. Adopting these intrusion detection models in the cloud layer, as shown in Fig. 7, can provide a warning to the system administrator when such attacks occur, which can verify early warnings (if they exist) from the EC nodes in the gateway layer. Collecting enough and meaningful data is very critical for AI techniques. This is considered a challenging step to reduce the error rate with these techniques. The cloud can detect any compromise by keeping logs of the presence of the connected gateways or ECs. It can also find alternative routes to IoMT sensors by providing a backup gateway in case of attacks, breaks, or loss of the original gateway.

## IX. CONCLUSIONS AND FUTURE WORK

Due to the demand for using IoMT sensors to reduce healthcare spending and provide better care for patients, securing these devices has become extremely important. However, IoMT sensors tend to have constrained resources, and some that are already implanted require external devices to secure them. In this paper, an overview of the security requirements, state-of-the-art security techniques, and new types of attacks were discussed. Since no one technique can satisfy these systems' security requirements and mitigate most of the attacks, we propose a framework that uses a combination of these techniques to meet all security requirements. This framework covers all data and device security stages, starting from data collection to data storage and sharing.

Since the framework proposed in this paper has some challenges, there is a need to design a system that can support a remotely secure initial setup and alternative access method. These methods can be used in case one of the factors is lost and in cases of emergencies.

**Author Biographies:**

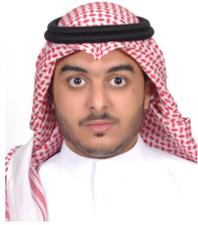

**Ali Ghubaish** (Graduate Student Member, IEEE) received the B.S. degree (Hons.) in computer engineering (minor in networking) from Prince Sattam Bin Abdulaziz University, AlKharj, Saudi Arabia, in 2013, and the M.S. degree in computer engineering from Washington University in St. Louis, MO, USA, in 2017, where he is currently pursuing the Ph.D. degree in computer engineering.
From 2013 to 2014, he worked as a Teaching Assistant at Prince Sattam Bin Abdulaziz University. Since 2018, he has been working as a Graduate Research Assistant at Washington University in St. Louis. His research interests include network and system security, the Internet of Things, the Internet of Medical Things, healthcare systems, and unmanned aerial vehicles (UAVs) communications.

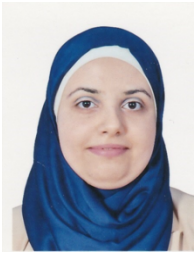

**Tara Salman** (Graduate Student Member, IEEE) received the B.S. degree in computer engineering and the M.S. degree in computing (networking minor) from Qatar University, Doha, Qatar, in 2012 and 2015, respectively. She is currently pursuing the Ph.D. degree in computer engineering at Washington University in St. Louis, MO, USA.
From 2012 to 2015, she worked as a Research Assistant with Qatar University on a National Priorities Research Program (NPRP) funded project, targeting physical layer security. She has been working as a Graduate Research Assistant with Washington University in St. Louis since 2015. She is an author of one book chapter and many international conferences and journals. Her research interests include network security, distributed systems, the Internet of Things, and financial technologies.

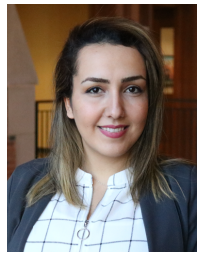

**Maede Zolanvari** (Graduate Student Member, IEEE) received the B.S. and M.S. degrees in electrical and computer engineering from Shiraz University, Shiraz, Iran, in 2012 and 2015, respectively. She is currently a Ph.D. candidate in Computer Science and Engineering at Washington University, St. Louis, MO, USA. From 2012 through 2015, her research was on performance improvement of communication networks, focusing on OFDM systems. Since 2015, she has been working as a graduate research assistant at Washington University. Her current research focuses on utilizing machine learning and deep learning for network security of the Industrial Internet of Things. Her research interests include the Internet of Things, machine learning, cyber-security, secure computer networks, and wireless communications.

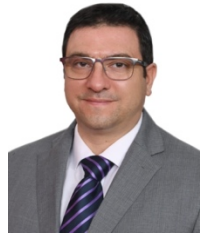

**Devrim Unal** (Senior Member, IEEE) is a Research Assistant Professor of Cyber Security at the KINDI Center for Computing Research, College of Engineering, Qatar University. He obtained his M.Sc. degree in Telematics from Sheffield University, UK, and Ph.D. degree in Computer Engineering from Bogazici University, Turkey, in 1998 and 2011, respectively. Dr. Unal's research interests include cyber-physical systems and IoT security, wireless security, artificial intelligence, and next-generation networks.

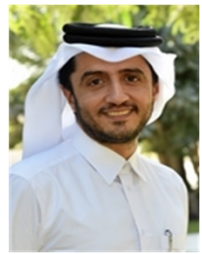

**Abdulla Al-Ali** (Member, IEEE) obtained his M.S. degree in Software Design Engineering and Ph.D. degree in Computer Engineering from Northeastern University in Boston, MA, USA in 2008 and 2014, respectively. He is an active researcher in cognitive radios for smart cities and vehicular ad-hoc networks (VANETs). Dr. Abdulla is currently head of the Computer Science and Engineering department at the College of Engineering in Qatar University.

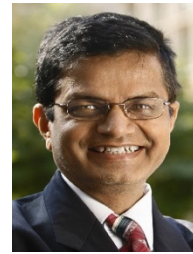

**Raj Jain** (Life Fellow, IEEE) received his B.E. degree in electrical engineering in 1972 from APS University, Rewa, India, an M.E. degree in Automation in 1974 from Indian Institute of Science, Bangalore, India, and a Ph. D. degree in Applied Maths (computer science) in 1978 from Harvard University, Cambridge Massachusetts, USA. He is currently the Barbara J. and Jerome R. Cox, Jr., Professor of computer science and engineering with Washington University in St Louis. He is a Fellow of the ACM and AAAS. He was one of the Co-founders of Nayna Networks, Inc., a next-generation telecommunications systems company in San Jose, CA, USA. He was a Senior Consulting Engineer at Digital Equipment Corporation in Littleton, Mass, and then a Professor of computer and information sciences, The Ohio State University in Columbus, Columbus, OH, USA. He is a recipient of the 2018 James B. Eads Award from St. Louis Academy of Science, the 2017 ACM SIGCOMM Life-Time Achievement Award, the 2015 A. A. Michelson Award from Computer Measurement Group. He ranks among the Most Cited Authors in Computer Science. He is the author of the Art of Computer Systems Performance Analysis, which won the 1991 ''Best-Advanced How-to Book, Systems'' award from the Computer Press Association.